\documentclass[12pt]{article}
\usepackage{tikz}
\usepackage{float}

\usepackage[utf8]{inputenc}
\usepackage{amsmath, amssymb}
\usepackage{graphicx}
\usepackage{hyperref}
\usepackage{caption}
\usepackage{geometry}
\usepackage[backend=biber,style=numeric-comp,sorting=none]{biblatex}
\usepackage{soul}
\usepackage{xcolor}
\sethlcolor{red}

\makeatletter
\renewcommand\maketitle{%
  \begin{center}
    {\LARGE \@title \par}%
    \vskip 1em
    {\large \@author \par}%
    \vskip 1em
    {\normalsize * These authors contributed equally to this work.\par}
  \end{center}
  \vskip 1em
  \@thanks 
}
\makeatother

\title{A Generative Model for Disentangling Galaxy Photometric Parameters}

\author{
  Keen Leung\textsuperscript{1,*}, 
  \thanks{Email: keen.l.bonnie@gmail.com}
  Colen Yan\textsuperscript{2,*},
  \thanks{Email: colenhcyan@gmail.com}
  Jun Yin\textsuperscript{3,*},
  \thanks{Email: junyin1313@gmail.com}\\
  \textsuperscript{1}Kristin School, Albany, Auckland, New Zealand \\
  \textsuperscript{2}Pinehurst School, Albany, Auckland, New Zealand \\
  \textsuperscript{3}Altas Science, Princeton, New Jersey, USA 
}

\date{February 2025}

\begin{document}

\maketitle

\begin{abstract}
Ongoing and future photometric surveys will produce unprecedented volumes of galaxy images, necessitating robust, efficient methods for deriving galaxy morphological parameters at scale. Traditional approaches, such as parametric light-profile fitting, offer valuable insights but become computationally expensive when applied to billions of sources. In this work, we propose a Conditional AutoEncoder (CAE) framework to simultaneously model and characterize galaxy morphology. Our CAE is trained on a suite of realistic mock galaxy images generated via \texttt{GalSim}, encompassing a broad range of galaxy types, photometric and morphological parameters (e.g., flux, half-light radius, S\'ersic index, ellipticity), and observational conditions. By encoding each galaxy image into a low-dimensional latent representation conditioned on key parameters, our model effectively recovers these morphological features in a disentangled manner, while also reconstructing the original image. The results demonstrate that the CAE approach can accurately and efficiently infer complex structural properties, offering a powerful alternative to existing methods. 
\end{abstract}

\section{Introduction}

Over the past two decades, astronomical photometric surveys have expanded our view of the universe, collecting vast datasets that contain billions of galaxies. Projects such as the Sloan Digital Sky Survey (SDSS), the Dark Energy Survey (DES), and upcoming projects such as Vera C. Rubin Observatory’s Legacy Survey of Space and Time (LSST) and the Nancy Grace Roman Space Telescope have made quality photometric observations available on a previously-unseen scale~\cite{york2000sdss, abbott2018des,ivezic2019lsst}. These large catalogs enable comprehensive studies of galaxy formation and evolution, including analyses of galaxy morphology and structure. Galaxy morphology, in particular, provides valuable insights into a galaxy’s formation history and evolutionary pathway~\cite{bamford2009galaxy, conselice2014evolution}. However, robustly deriving morphological parameters such as brightness, size, and shape across billions of sources is a formidable challenge, both in terms of computational resources and methodological complexity.

Traditional parametric modeling techniques (e.g., fitting S\'ersic profiles) remain popular for extracting structural properties from galaxy images~\cite{peng2002galfit}, but they are computationally intensive and often rely on restrictive assumptions about the galaxy’s light profile. As surveys grow ever larger, there is a clear need for more scalable methods that can accurately model galaxy morphologies under realistic observational conditions. In recent years, machine learning has emerged as a powerful toolkit for handling high-dimensional, noisy astrophysical data~\cite{fluke2020survey, baron2019machine, banerji2010galaxy, barchi2020machine, dawes10review}. In particular, generative models have garnered interest for their ability to learn complex data distributions in an unsupervised or semi-supervised manner.

Among generative approaches, autoencoders stand out as a flexible framework capable of learning low-dimensional latent representations that capture the essential features of galaxy images~\cite{yin2022conditional, scourfield2023noising,mirzoyan2025enhancing, buonanno2020deep, bretonniere2012euclid, lanusse2020deep}. When implemented in a conditional setting—i.e., incorporating known labels or physical parameters—these models can help disentangle different aspects of galaxy structure (e.g., flux, shape, orientation) in a principled way. By mapping high-dimensional inputs to a compressed latent space and back, a conditional autoencoder can perform both data compression and reconstruction while preserving physically meaningful properties. This approach thereby offers a promising avenue for addressing the computational challenges of galaxy morphology analysis in modern photometric surveys.  

In this work, we investigate the use of a conditional autoencoder for galaxy photometric parameter estimation. While machine learning has previously been applied to galaxy morphology in various contexts~\cite{yin2022conditional, scourfield2023noising, banerji2010galaxy, buonanno2020deep}, our approach differs in several important ways. First, we adopt a significantly simpler neural network architecture, which enhances the interpretability and disentanglement of the learned features. This emphasis on an interpretable latent space is aligned with recent work by Csizi et al., who also employed compressed and interpretable latent representations as the foundation for a deep generative galaxy morphology model~\cite{csizi2025euclid}. Second, we conduct a systematic comparison with traditional statistical methods to quantify the complexity of the task, thereby justifying the use of machine learning tools.

We begin by generating a suite of realistic mock galaxy images that simulate a variety of structural properties and observational effects in Section~\S\ref{sec:data}. We then discuss our autoencoders architecture and loss function design in Section~\S\ref{sec:CAE}. Finally, we present our results in Section~\S\ref{sec:results} and discuss the implications for large-scale galaxy surveys, highlighting both the advantages and remaining challenges of the generative approach.

\section{Data}
\label{sec:data}

We generate our mock galaxy images using the open-source \texttt{GalSim} package~\cite{galsim_package}, which is designed to simulate realistic images of astronomical objects, particularly galaxies. \texttt{GalSim} provides flexible routines to model galaxy light profiles, instrumental point spread functions (PSFs), and noise properties. This flexibility makes it well suited for producing training and validation sets for machine learning tasks in astronomy.

\subsection{Galaxy Model: S\'ersic Profile}

For each mock galaxy, we adopt a S\'ersic profile as the underlying light distribution. The S\'ersic profile is a parametric model commonly used in astrophysics to describe the radial surface brightness of galaxies~\cite{sersic1968atlas}. It is given by:
\begin{equation}
    I(r) = I_e \exp \Biggl\{ 
        -b_n \Bigl[ \Bigl(\frac{r}{r_e}\Bigr)^{1/n} - 1 \Bigr] 
    \Biggr\},
    \label{eq:sersic}
\end{equation}
where \(r\) is the radial distance from the galaxy center, \(r_e\) is the effective (or half-light) radius, \(I_e\) is the intensity at \(r_e\), \(n\) is the S\'ersic index controlling the concentration of the profile, and \(b_n\) is a constant defined in terms of \(n\) such that \(r_e\) encloses half of the total light. In our simulations, \(n\) varies between 0.5 and 4.0, covering disk-like to bulge-dominated morphologies.

\begin{table}[ht]
\centering
\caption{Parameter ranges for galaxy simulation with \texttt{GalSim}.}
\label{tab:galsim_parameters}
\begin{tabular}{lccc}
\hline
\textbf{Parameter} & \textbf{Symbol} & \textbf{Range} & \textbf{Description} \\
\hline
Flux & $F$ & 500 -- 50{,}000  & Total integrated flux (in ADU) \\
Half-light radius & $r_{1/2}$ & 0.3 -- 5.0 arcsec & Characteristic size of the galaxy \\
S\'ersic index & $n$ & 0.5 -- 4.0 & Concentration of light profile \\
Ellipticity & $e$ & 0.0 -- 0.6 & Galaxy axis ratio and orientation \\
Position angle & $\theta$ & 0 -- 180 degrees & Rotation of the galaxy major axis \\
$x$-coordinate & $x_0$ & 0 -- 64 pixels & Galaxy center (horizontal) \\
$y$-coordinate & $y_0$ & 0 -- 64 pixels & Galaxy center (vertical) \\
\hline
\end{tabular}
\end{table}

The parameter ranges in Table~\ref{tab:galsim_parameters}  are chosen to be broad yet physically reasonable for typical ground-based observations, while avoiding extreme galaxy configurations that would introduce strong degeneracies or numerical instabilities. For example, we restrict ellipticities to $e \leq 0.6$ to exclude highly elongated systems that are less common and more difficult to model reliably at modest resolution, and we adopt Sérsic indices in the range $0.5$-$4.0$ to cover the dominant morphological classes without entering regimes where the profiles become extremely peaked. All parameters are sampled from uniform distributions, which allows us to construct a controlled dataset that does not encode astrophysical populations priors. This ensures that the conditional autoencoder is evaluated on its ability to learn the parameter-morphology relationships themselves, independent of any underlying distribution of galaxy properties.

\subsection{Point Spread Function and Noise Model}

To mimic observational conditions, each galaxy model is convolved with a point-spread function. The PSF describes how light from a point source is distributed on the detector, encapsulating atmospheric or instrumental effects. In our setup, we use a Gaussian PSF with a typical full width at half maximum (FWHM) of \(\approx 0.7''\), representative of ground-based seeing conditions. 

After convolving each galaxy with the PSF, we add a sky background of 100 ADU per pixel and Poisson noise to emulate photon-counting statistics, as well as a small Gaussian read noise to approximate realistic detector behavior. These steps collectively generate our mock training and validation dataset for the conditional autoencoder, ensuring exposure to a variety of galaxy morphologies and observational conditions.

\section{Methods}
\label{sec:CAE}
\subsection{Convolutional Autoencoder}

An autoencoder is an unsupervised neural network that learns a compact latent representation of an input $\mathbf{x}\in\mathbb{R}^{D}$ through two mappings: an encoder $f_{\theta}: \mathbf{x}\mapsto\mathbf{z}\in\mathbb{R}^{d}$ and a decoder $g_{\phi}: \mathbf{z}\mapsto\widetilde{\mathbf{x}}$, with $d\ll D$~\cite{hinton2006reducing}.  The parameters $(\theta,\phi)$ are optimized to minimize the difference between the input and reconstructed output, thereby forcing the bottleneck $\mathbf{z}$ to retain the information most relevant for accurate synthesis~\cite{goodfellow2016deep}. Because of the nonlinear transformations in encoder–decoder, autoencoders learn to capture data distributions better than linear techniques such as principal–component analysis when the underlying distribution is complex. Furthermore, the latent representations can be efficiently trained to disentangle different generative factors of the data~\cite{kingma2019introduction, burgess2018understanding, sikka2019closer}. Due to the desirable properties of latent interpretability and disentanglement, autoencoders have garnered increasing attention in the physical sciences~\cite{carleo2019machine,yang2023detecting,luchnikov2019variational,gheller2022convolutional,wetzel2017unsupervised,bjerrum2018improving,romero2017quantum,zhong2021machine,zhong2020learning,zhong2023non, ochiai2023variational, yang2022learning, finke2021autoencoders}. 

Fully-connected autoencoders are prone to overfitting and ignoring spatial locality - limitations that are acute for high-resolution images~\cite{zhang2018better}. Convolutional autoencoders, on the other hand, alleviate these issues by replacing dense layers with weight-sharing kernels that slide across the input; pooling layers down-sample spatial dimensions in the encoder, while transposed convolutions or nearest-neighbour up-sampling restore the original resolution in the decoder~\cite{dumoulin2016guide}. With locality build-in to the shared convolutional kernels, convolutional autoencoders are more parameter efficient and achieve superior performance on image data, making them a popular choice for scientific image processing (e.g.~\cite{chen2017deep, fukami2020convolutional, farina2020searching, yin2021active}).

For the analysis below, we employ a hybrid design: convolutional layers extract hierarchical features in both encoder and decoder, whereas a compact fully connected bottleneck aggregates global context.  This configuration combines the spatial awareness of convolutional layers and the interpretability of dense layers, allowing for a robust analysis of galaxy photometry features. 

Crucially, when we subsequently analyze the relationship between latent dimensions and physical properties, we refer exclusively to dimensions within the learned latent bottleneck prior to concatenation, not the augmented vector. This distinction is important because the concatenated parameters are directly embedded into the decoder input and would exhibit perfect correlation with their corresponding physical properties. Our analysis instead probes whether the learning process has caused individual dimensions of the original learned latent space to encode meaningful variations, independent of the concatenated constraints.

\begin{figure}[!t]
    \centering
    \includegraphics[width=0.7\linewidth]{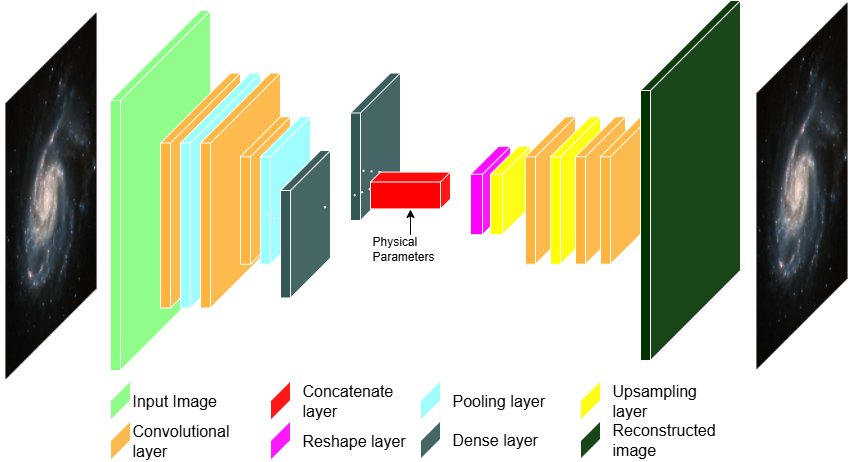}
    \caption{Representation of the conditional autoencoder, taking an input image of a galaxy to process through the encoder, latent space and a decoder to reconstruct the output image for photometric analysis. The latent space is supervised to learn the physical parameters of the galaxy image. Detailed parameters can be found in the Appendix.}
        \label{fig:CAE}
\end{figure}

\subsubsection{Conditional autoencoder}
\label{sec:loss_function}
A Conditional AutoEncoder (CAE) is a autoencoder conditioned on some data parameters during the decoding stage. We used a CAE to both reconstruct the galaxy images and learn the physical parameters in its latent space.

The loss function of our CAE consists of three parts: reconstruction loss, physical parameter loss, and regularization loss:
\begin{equation}
\label{eq:L_total}
    L_{\mathrm{total}}
= L_{\mathrm{rec}} + \alpha \, L_{\mathrm{phy}} + \beta  L_{\mathrm{reg}}.
\end{equation}

Overall, our loss function measures how well the model's predictions match the input data and how well the latent space captures the physical parameters. In particular, the loss function terms are as follows:

\begin{itemize}
    \item Binary cross-entropy (BCE), which treats each normalised pixel as a Bernoulli variable and penalises reconstruction errors by:
    \begin{equation}
    L_{\text{rec}}
    \;=\;
    \frac{1}{BHW}\,
    \sum_{b=1}^{B}\;
    \sum_{p=1}^{HW}
    \Bigl[
      -\,x_{bp}\,\log\!\bigl(\hat{x}_{bp}\bigr)
      \;-\;
      \bigl(1-x_{bp}\bigr)\,
      \log\!\bigl(1-\hat{x}_{bp}\bigr)
    \Bigr],        
    \end{equation}
\end{itemize}
where \(B = 64\) is the batch size, \(H \times W = 64 \times 64\) is the image resolution, \(x_{bp} \in [0,1]\) is the target pixel value, \(\hat{x}_{bp} \in [0,1]\) is the network’s prediction.

\begin{figure*}[!t]
    \centering
    \includegraphics[width=1\linewidth]{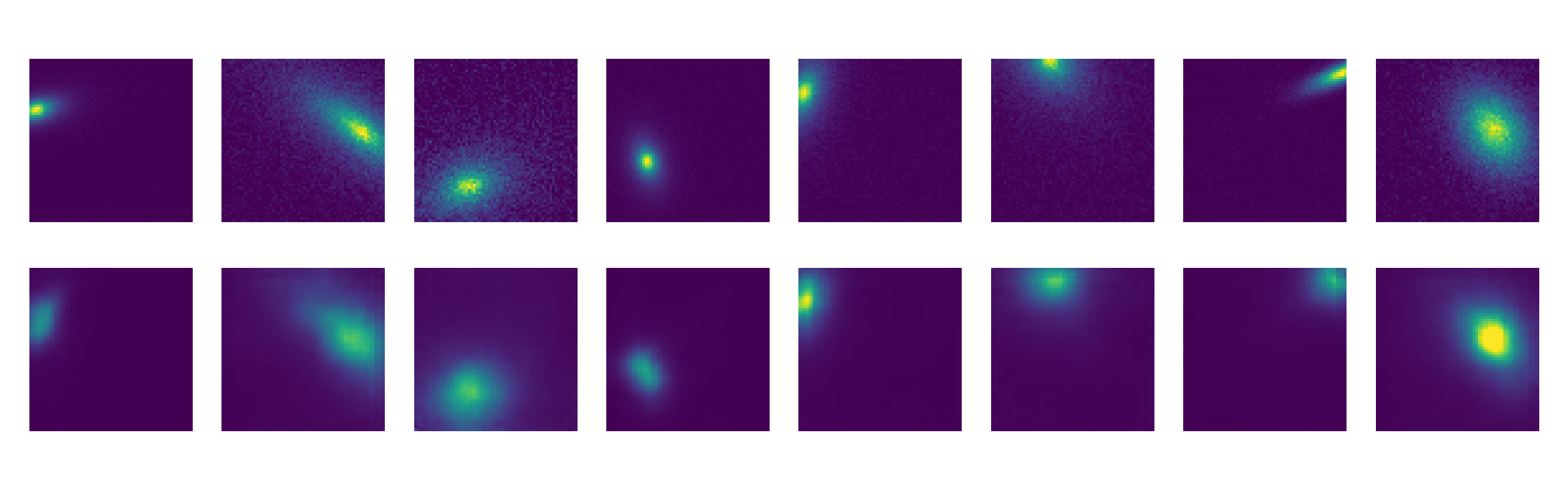}
    \caption{Sample reconstructions from the autoencoder. Top row: input galaxy images. Bottom row: corresponding reconstructions, normalized to the same range as the input on top. }
    \label{fig:reconstruction}
\end{figure*}

\begin{itemize}
    \item Physical parameter loss, which penalises the discrepancy between the true physical parameters $z_{\text{phy}} \in \mathbb{R}^{n_{\text{phy}}}$ and the predicted ones $\hat{z}_{\text{phy}}$:
    \begin{equation}
            L_{\mathrm{phy}}
        = \frac{1}{n_{\text{phy}}} \sum_{i=1}^{n_{\text{phy}}} (z^{\text{phy}}_i - \hat{z}^{\text{phy}}_i)^2
    \end{equation}
\end{itemize}
where $n_{\text{phy}}=2$ is the number of physical parameters we supervise in the latent space.

\begin{itemize}
    \item L-1 regularization on the rest of the latent space (excluding the physical ones)
    \begin{equation}
        L_{\mathrm{reg}} = \frac{1}{d-n_{\text{phy}}} \sum_{i=1}^{d-n_{\text{phy}}} \left| z^{\text{rest}}_i - \hat{z}^{\text{rest}}_i \right|,
    \end{equation}
\end{itemize}
where $d=64$ is the total dimension of the latent space. In the context of representation learning, applying L1 regularization to the latent parameters of an autoencoder can encourage sparsity in the learned latent space. Regularization drives \( z^{\text{rest}} \) toward zero, promoting compact and interpretable representations~\cite{ng2004feature, tibshirani1996lasso}. We choose not to regularize the physical latent space similar to previous proposals~\cite{zhao2023sparse, schar2005heatwaves}, and only regularize the residual non-essential latent parameters \( z^{\text{rest}} \).

The architecture of the CAE is presented in Figure~\ref{fig:CAE}. Composing our encoder is a sequence of convolutional layers followed by max-pooling layers to reduce dimensionality, and a flatten layer followed by dense layer at the latent space. At the latent space, $z^{\text{phy}}$ is a set of latent dimensions which are isolate the known values of specific physical parameters (e.g. flux, half-light radius). $z^{\text{rest}}$ is encoded from the input images and captures non-specified morphological data (e.g. S\'ersic index). They are subsequently concatenated into a single dense layer, $z$. The decoder is composed of de-convolutional layers and upsampling layers to reconstruct the image.

\section{Results}
\label{sec:results}

To evaluate whether our autoencoder is capable of learning meaningful representations of galaxy images, we first assessed its ability to reconstruct diverse examples from \texttt{GalSim}. We prepared 10000 simulated galaxy samples, split into a train-to-test ratio of 10:1. Each input image was generated by randomly sampling a realistic range of galaxy parameters, including total flux, half-light radius (arcsec), Sérsic index, ellipticity, position angle (degrees), and central coordinates $(x, y)$ in pixel space (see Table~\ref{tab:galsim_parameters}). Figure~\ref{fig:reconstruction} shows a selection of the original input (top row) along with their reconstructions (bottom row). The reconstructions closely preserve key visual and structural features of galaxies—such as shape, orientation, and brightness distribution—demonstrating that the model has learned an effective representation of the input images.

\begin{figure}[!t]
    \centering
    \includegraphics[width=0.70\linewidth]{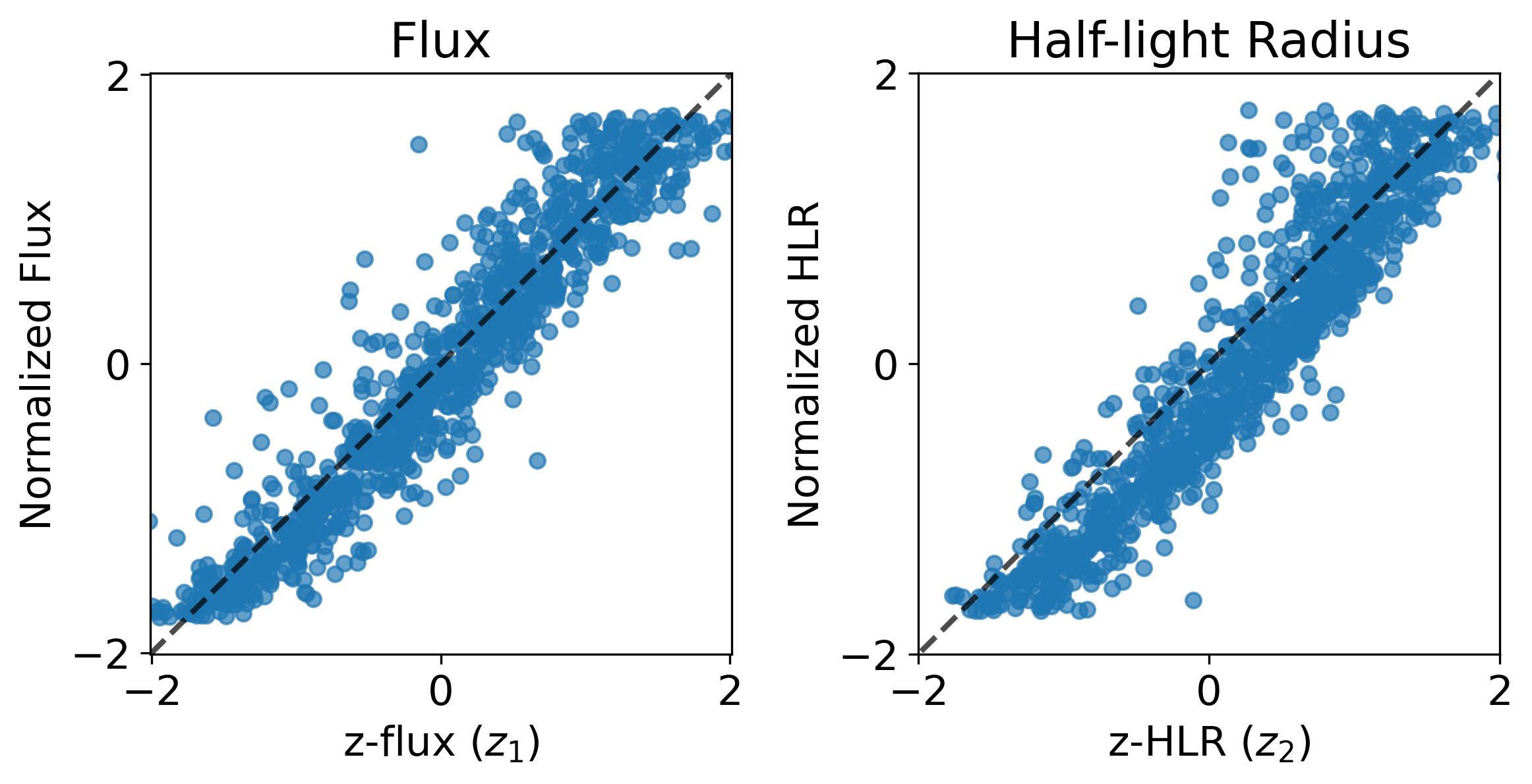}
    \caption{Correlation between physical parameters and latent dimensions. Left: flux vs $z_1$; right: half-light radius vs $z_2$. Each point represents a galaxy image sample.}
    \label{fig:latent_corr}
\end{figure}

To evaluate whether our autoencoder learns physically meaningful representations of galaxy properties, we focused on two key parameters: total flux and half-light radius. These two parameters were specifically chosen as total flux is a global property which must be inferred as the limited image means that the total flux is cut off, hence being a metric for how well our model can predict a value from limited data, and half-light radius is key for morphological scale length. We choose these two most significant parameter of variations as an proof-of-concept of our method, and the framework can be extended to other parameters. Using controlled datasets generated with \texttt{GalSim}, we independently varied each parameter across a fixed range while holding the others constant. When encoding these images, we observed strong, nearly linear correlations between total flux and the first latent dimension $z_1$, as shown in Figure \ref{fig:latent_corr}. However, qualitative observation of the half-light radius and the second latent dimension $z_2$ suggests that there is systematic error where the second latent dimension consistently overestimates normalised HLR. Despite this, it still shows a significant correlation, with the $R^2$ value of the graph being 0.901. In all, these results suggest that the autoencoder has learned to disentangle and internally represent both flux and size as distinct, interpretable axes in its latent space.

\begin{figure}
    \centering
    \includegraphics[width=0.35\linewidth]{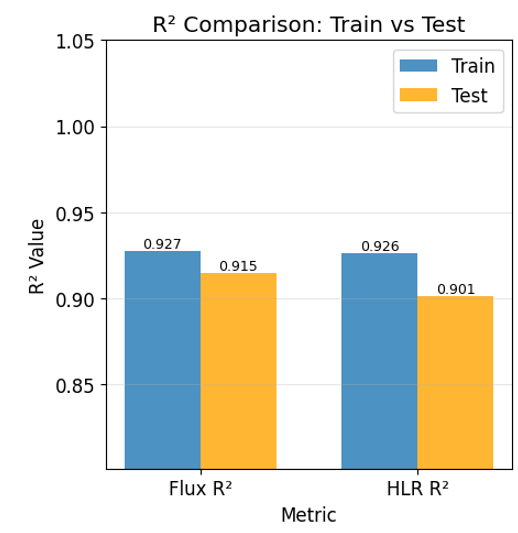}
    \caption{Comparison of $R^2$ between input and output parameters for train and test datasets.}
    \label{fig:R^2compar}
\end{figure}
\begin{figure}[!t]
  \makebox[\textwidth]{\hspace*{-0.01\textwidth}%
    \includegraphics[width=1.05\linewidth]{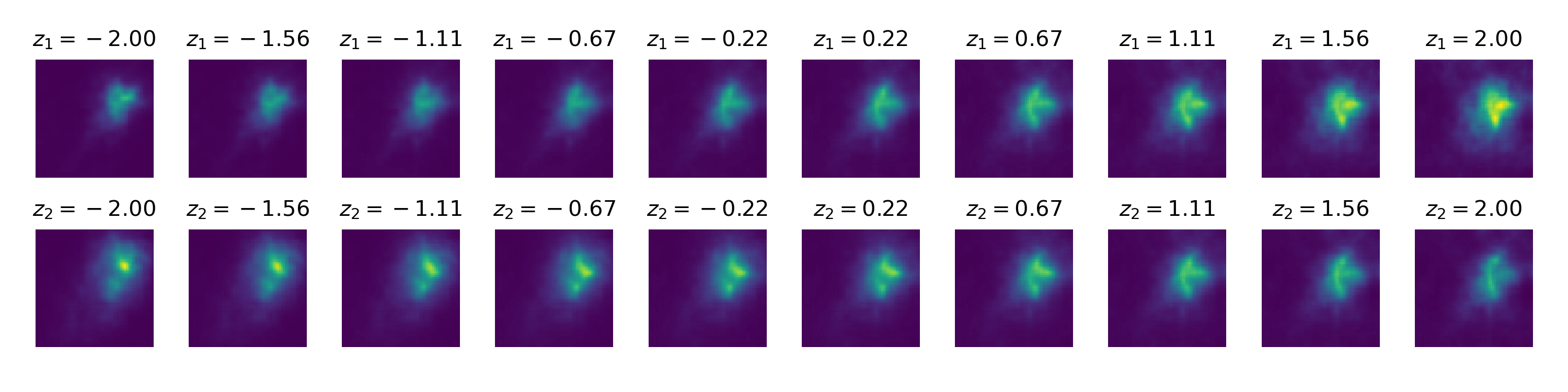}}
    \caption{Effect of varying individual latent dimensions. Top row: increasing $z_1$ (flux) causes brightness to increase while structure remains fixed. Bottom row: increasing $z_2$ (half-light radius) leads to a visibly larger galaxy size. All other latent variables are held constant; both rows originate from the same base galaxy.}
    \label{fig:dual_sweep}
\end{figure}

To test whether individual latent dimensions within $z^{\text{rest}}$ actively control specific physical properties, we performed independent latent space sweeps along $z_1$ and $z_2$, corresponding to flux and half-light radius, respectively. Starting from a fixed latent vector representing a sample galaxy, we varied one component at a time over a uniform range while keeping all others constant. As shown in Figure \ref{fig:dual_sweep}, increasing $z_1$ produces a sequence of reconstructions with progressively greater brightness, while morphology remains unchanged. In contrast, varying $z_2$ leads to visibly larger galaxy profiles, indicating an increase in half-light radius without affecting overall brightness. These results confirm that $z_1$ and $z_2$ function as generative axes for flux and size, respectively, demonstrating that the autoencoder’s latent space is not only compact and disentangled, but also interpretable and responsive to meaningful astrophysical variations.

Furthermore, for both metrics, we divided our train and test datasets to get their individual $R^2$ values, shown in Figure \ref{fig:R^2compar}. Given the difference between $R^2$ values, it is reasonable to say that there is a moderate but acceptable performance drop in the test dataset.

\begin{figure}[!t]
  \makebox[\textwidth]{\hspace*{-0.01\textwidth}%
    \includegraphics[width=0.7\linewidth]{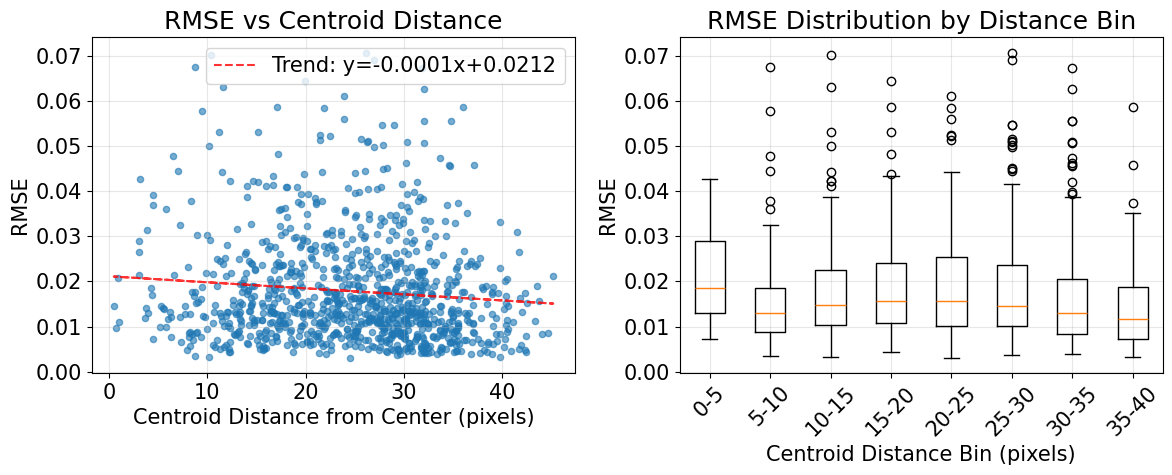}}
    \caption{Variation of RMSE by centroid distance. Left: RMSE vs centroid distance; right: RMSE distribution by distance bin.}
    \label{fig:centroid}
\end{figure}

It should be acknowledged that, in real scenarios, the centroid position of a galaxy can be detected by image pre-processing, and in the current study the variable centroid position is chosen to improve the robustness of our model. As shown in Figure \ref{fig:centroid}, a larger centroid distance from the center does not tend to increase RMSE, suggesting our model is robust at larger distances.

For quantitative metrics examining the efficacy of the model, SSIM (Structural Similarity Index) is effective for evaluation, as it captures the success of preserving morphological information. The average SSIM between the reconstructed and original datasets is 0.788, which is an acceptable value~\cite{wang2004SSIM}, indicating that the general structure is preserved but fine details are compromised.

To further benchmark our model, we decided to test a further 10000 GalSim-generated images with more realistic parametric distributions as opposed to uniformly sampling from our parameter ranges, outlined in Table \ref{tab:cosmos_parameters}. In this realistic simulation, flux was chosen to COSMOS clipped normal distribution with fainter galaxies~\cite{leauthaud2007weak}. Half-light radius' distribution is chosen to be a log-normal distribution with a larger number of smaller galaxy sizes ~\cite{sargent2007evolution,koekemoer2007cosmos}, and the S\'ersic index is chosen to be bimodal normal distribution~\cite{sargent2007evolution}.  The ellipticity is chosen to be Rayleigh-like~\cite{leauthaud2007weak,mandelbaum2014third}. Lastly, the centroids are positioned in the center.

\begin{table}[ht]
\centering
\caption{Parameter ranges for GalSim with COSMOS distributions and ranges.}
\label{tab:cosmos_parameters}
\begin{tabular}{|p{3cm}|p{3.5cm}|p{3cm}|p{4.5cm}|}
\hline
\textbf{Parameter} & \textbf{Range (with units)} & \textbf{Distribution Type} & \textbf{Distribution Factors} \\
\hline
Flux & $500 - 50{,}000$ ADU & Normal (clipped) & $\mu = 23.0$ mag, $\sigma = 1.5$, $\mathrm{ZP} = 25.94$ \\
\hline
Half-light radius & $1.0 - 12.0$ pix ($0.05 - 0.60$ arcsec) & Log-normal (scaled) & $\mu_{\log} = 1.50$, $\sigma_{\log} = 0.4$, slope $= -0.15$ \\
\hline
Sersic index & $0.5 - 6.0$ & Bimodal normal & Disk: $\mu = 1.2$, $\sigma = 0.3$ (70\%); Bulge: $\mu = 3.8$, $\sigma = 0.5$ (30\%) \\
\hline
Ellipticity & $0.0 - 0.85$ & Rayleigh (conditional) & scale $= 0.25$; suppressed for $n > 3.0$ \\
\hline
Position angle & $0° - 180°$ & Uniform & N/A \\
\hline
\end{tabular}
\end{table}

From the more realistically distributed simulations, we found that the average SSIM between reconstructed and original datasets is 0.955, which is a very high value in comparison with the SSIM of 0.788 of the uniformly distributed dataset. This indicates that the conditions associated with actual surveys (be it the smaller half-light radius or fainter flux) is more favorable for reconstruction, discounting uncontrolled variables such as cosmic rays that interfere with real data.

\subsection{Comparative Analysis: PCA vs. CAE}

Principal Component Analysis (PCA) is a classical linear technique for dimensionality reduction, frequently employed to simplify high-dimensional datasets by projecting them onto a lower-dimensional subspace that captures the directions of maximal variance~\cite{jolliffe2016pca}. In particular, PCA is equivalent to a linear autoencoder~\cite{baldi1989pca}. We decode the latent factors discovered by PCA with a linear decoder to create a baseline and benchmark our model.

To evaluate the capacity of linear methods to capture physical parameters from galaxy images, we performed PCA on the same set of data using $d=64$ principal components, matching the number of total CAE latent dimensions. We first applied PCA in a constrained setting: we performed linear regression on each of the 64 principal components for the flux and half-light radius to predict those physical parameters, and computed the corresponding $R^2$. As a comparison with the physical latent parameter, we picked the single best principal component for each target variable - defined as the component that yielded the highest $R^2$. In this setting, performance was negligible, with $R^2$ values near zero for both the flux and the half-light radius. This result indicates that no individual linear projection can meaningfully capture these parameters, which illustrates the difficulty of the task.

\begin{figure}
    \centering
    \includegraphics[width=0.65\linewidth]{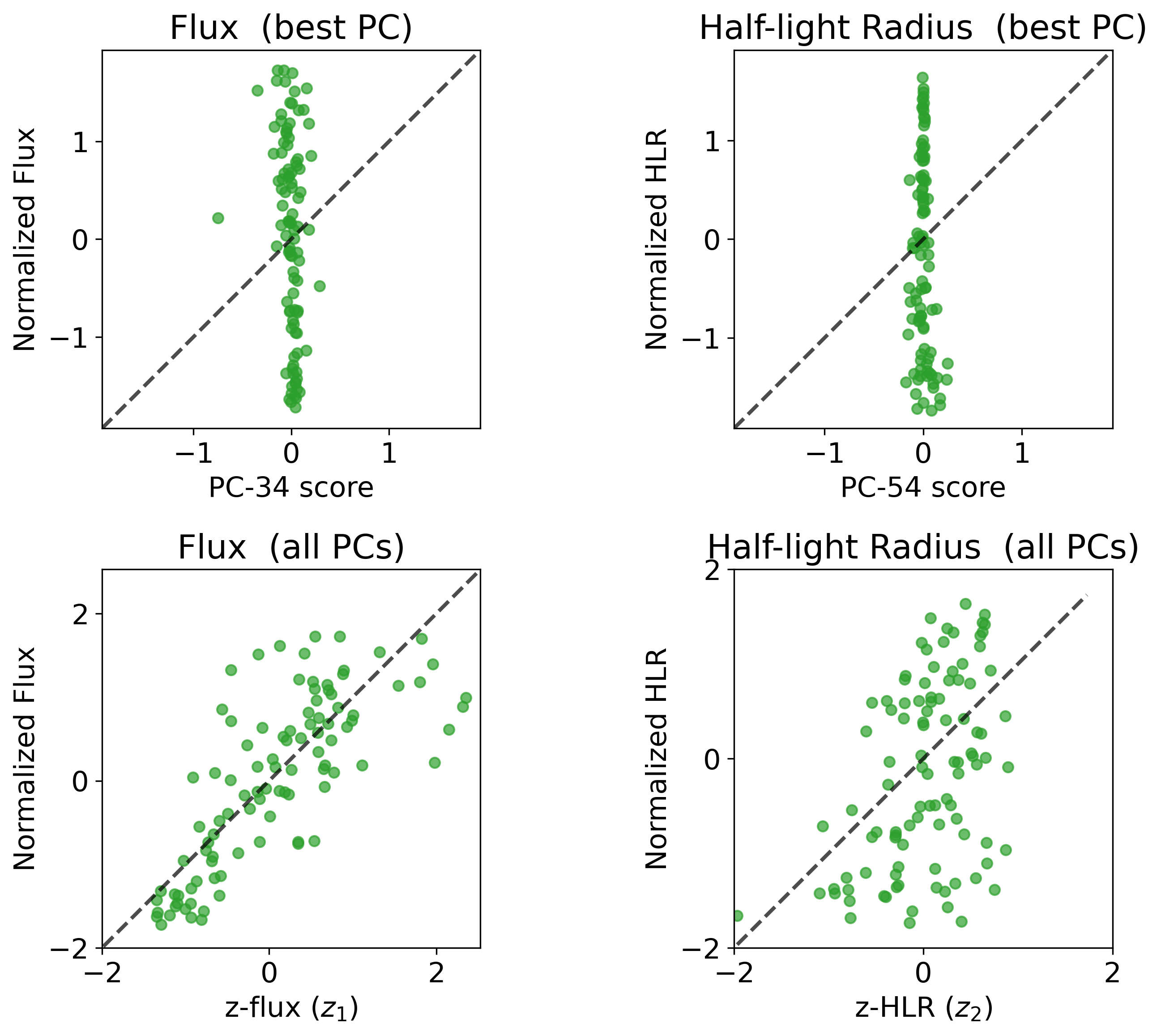}
    \caption{Correlation between physical parameters and latent dimensions. Top row: using only the best principal component to predict the corresponding physical parameter. Bottom row: using all the principal components to predict the physical parameter. Left: flux; right: half-light radius. Each point represents a galaxy image sample.}
    \label{fig:PCA_comp}
\end{figure}

We then allowed PCA to use all 64 principal components, vastly out-numbering the two-component physical latent space we used in the CAE, and used it for linear regression on the physical parameters. Predictive performance improved, but remained modest, with $R^2 = 0.467$ for flux ($z_1$) and $R^2 = 0.137$ for the half-light radius ($z_2$). In contrast, our autoencoder achieved significantly higher scores, $R^2 = 0.775$ and $R^2 = 0.722$, respectively, using only one latent variable per parameter. This demonstrates that the autoencoder learns compact, disentangled, and semantically meaningful representations that outperform PCA even under unfavorable conditions. A visual comparison of PCA performance across both dimensionalities is shown in Figure~\ref{fig:PCA_comp}. Through this comparison, it is apparent that the features discovered by linear decoders were insufficient as galaxy datasets are inherently nonlinear in nature. Therefore, methods such as CAE holds more promise for capturing galaxy photometries obtained from real observational data. 

\section{Discussion}

\paragraph{Summary.} This proof--of--concept study demonstrates that a conditional autoencoder (CAE) can recover physically meaningful latent variables from galaxy images while remaining computationally efficient.  In our experiments, we assign total flux and half-light radius using our $z_\mathrm{phy}$ layer to two dedicated physical latent coordinates, $z_{1}$ and $z_{2}$, respectively. The strong one--to--one relations in Figure \ref{fig:latent_corr} confirm that the supervision term in Eq.~\eqref{eq:L_total} successfully aligns these axes with their target parameters.  Manipulating each coordinate in isolation produces the expected response in image space—brightness varies smoothly with $z_{1}$, whereas the apparent size changes with $z_{2}$ (Figure~\ref{fig:dual_sweep}).  Such behavior indicates a disentangled latent geometry in which individual dimensions control single astrophysical factors of variation.

A comparison with the classical linear baseline further underlines the advantages of a non-linear, convolutional architecture.  Principal-component analysis (PCA) requires all 64 principal components to attain markedly lower predictive power than the CAE achieves with one coordinate per parameter (Figure~\ref{fig:PCA_comp}).  The gap implies that the manifold of realistic galaxy morphologies is strongly non-linear and is captured more faithfully by convolutional features than by global linear projections.

\paragraph{Limitations and future work.}  Several caveats must be addressed before the method can be deployed in production pipelines.  (i) Simulated data.  Our network is trained exclusively on idealised \texttt{GalSim} images; real survey data contain blended sources, cosmic rays, detector artefacts and rare morphologies that can degrade performance. Hence, an important extension would be to utilise real data from surveys such as \textit{Euclid} or LSST. Domain-adaptation strategies will be necessary. For example, concerning blending, a strategy than can be adopted is forcing the encoder to produce similar latent features for simulated and blended galaxies, making the decoder more robust to blending artifacts~\cite{brauer2025simulationssurveysdomainadaptation}  (ii) Limited parameter set.  We supervise only two parameters.  Extending the framework to ellipticity, Sérsic index, bulge--to--disc ratio or point--source contamination is feasible but will require a larger latent dimensionality and stronger regularization.  (iii) High--concentration profiles.  The CAE occasionally under--fits highly concentrated ($n\gtrsim3$) galaxies.  Incorporating a multi--scale or parametric likelihood loss may alleviate this. 

\paragraph{Implications.}  Ongoing surveys such as Rubin, \textit{Euclid} and the Nancy Grace Roman Space Telescope will image \emph{tens of billions} of galaxies.  Traditional maximum--likelihood profile--fitting codes cannot scale to that data volume.  Our results indicate that conditional autoencoders provide a viable alternative: they amortise the computational cost of fitting into a single forward pass, admit fast uncertainty estimation (e.g. via Monte--Carlo dropout) and yield a latent space in which astrophysical parameters are explicitly encoded.  We therefore expect deep generative models to play an increasingly central role in large--scale galaxy morphology studies in the coming decade.

\section*{Acknowledgements}
The authors would like to thank Weishun Zhong for suggesting this project and for providing guidance throughout the numerical experiments and the preparation of this manuscript.

\section*{Funding}
The authors do not have any funding information to disclose.

\appendix
\section*{Appendix}
\addcontentsline{toc}{section}{Appendix} 

\subsection*{Autoencoder Architecture}

The primary objective of our autoencoder architecture is to provide architecture which works well as a proof-of-concept for future use of conditional autoencoders.

The conditional autoencoder processes grayscale input images of dimensions $64 \times 64 \times 1$. The encoder employs hierarchical feature extraction  comprising two convolutional layers: one with 32 filters ($3\times3$ kernel size), ReLU activation, and same padding, followed by a max pooling layer with a $2\times2$ window size and same padding. The subsequent convolutional layer has 64 filters with identical kernel dimensions and activation, followed by equivalent a max pooling layer. This pipeline results in an encoded representation of size $16\times16\times64$.

The latent space has a bifurcated architecture, split into two dense layers: $z^{\text{phy}}$, dense linear layer with 2 units, and $z^{\text{rest}}$, dense 62 unit ReLU activation layer, incorporating L1 regularization. These are followed by concatenation to form a 64 dimensional latent vector.

The decoder module initiates through a dense layer projecting to 16,384 units with ReLU activation, then reshaped to $16\times16\times64$. Feature upscaling occurs via two consecutive upsampling-convolution layers, with each upsampling later applying $2\times2$ upsampling, each followed by convolutional layers with 64 and 32 filters respectively ($3\times3$ kernels, ReLU activation, same padding). The reconstruction concludes with an output convolutional layer with a $3\times3$ filter and sigmoid activation, producing the $64\times64\times1$ output.

Our CAE training is performed with a learning rate of $10^{-3}$ for 100 epochs using a batch size of 64. The loss weights in Eq.~\eqref{eq:L_total} are set to $\alpha = 1$ and $\beta = 10^{-4}$.

\printbibliography

\end{document}